\documentclass{article}
\usepackage{arxiv}

\usepackage[utf8]{inputenc} 
\usepackage[T1]{fontenc}    
\usepackage{hyperref}       
\usepackage{url}            
\usepackage{booktabs}       
\usepackage{amsfonts}       
\usepackage{nicefrac}       
\usepackage{microtype}      
\usepackage{lipsum}
\usepackage{float}
\usepackage{graphicx}
\usepackage{multirow}
\usepackage{array}
\usepackage{comment}
\usepackage{longtable}
\usepackage{ragged2e}
\usepackage{subfig}
\usepackage[table]{xcolor}  
\usepackage{geometry}
\usepackage{multicol}
\usepackage{amsmath}
\usepackage{pifont}
\title{PMI-DT: Leveraging Digital Twins and Machine Learning for Predictive Modeling and Inspection in Manufacturing}

\author{
  Chas Hamel \\
  Industrial and Systems Engineering\\
  University of Oklahoma\\
  Norman, Oklahoma-73069 \\
  \texttt{darren.c.hamel-1@ou.edu} \\
   \And
  Md Manjurul Ahsan \\
  Department of Industrial and Systems Engineering\\
  University of Oklahoma\\
  Norman, Oklahoma-73071\\
  \texttt{ahsan@ou.edu} 
   \And 
  Shivakumar Raman \\
  Department of Industrial and Systems Engineering\\
  University of Oklahoma\\
  Norman, Oklahoma-73071\\
  \texttt{raman@ou.edu} 
   \And
}

\begin{document}
\maketitle

\begin{abstract}
Over the years, Digital Twin (DT) has become popular in Advanced Manufacturing (AM) due to its ability to improve production efficiency and quality. By creating virtual replicas of physical assets, DTs help in real-time monitoring, develop predictive models, and improve operational performance. However, integrating data from physical systems into reliable predictive models, particularly in precision measurement and failure prevention, is often challenging and less explored. This study introduces a Predictive Maintenance and Inspection Digital Twin (PMI-DT) framework with a focus on precision measurement and predictive quality assurance using 3D-printed 1”-4 ACME bolt, CyberGage 360 vision inspection system, SolidWorks, and Microsoft Azure. During this approach, dimensional inspection data is combined with fatigue test results to create a model for detecting failures. Using Machine Learning (ML) —Random Forest and Decision Tree models—the proposed approaches were able to predict bolt failure with real-time data 100\% accurately. Our preliminary result shows Max Position (30\%) and Max Load (24\%) are the main factors that contribute to that failure. We expect the PMI-DT framework will reduce inspection time and improve predictive maintenance, ultimately giving manufacturers a practical way to boost product quality and reliability using DT in AM.

\end{abstract}


\keywords{Digital Twin \and Predictive Modeling \and Quality Control \and Machine Learning \and Advanced Manufacturing}

\section{Introduction}\label{sec1}

Digital Twin (DT) technology is a virtual model, or "twin," of a physical object that continuously updates with real-time data from its physical counterpart. This integration allows detailed monitoring, simulation, and predictive analysis, making DTs highly useful in manufacturing for managing product lifecycles, predictive maintenance, and optimizing processes through data drawn from sensors, machine learning models, and the Internet of Things (IoT)~\cite{rosen2015autonomy, tao2018digitaltwin}.

The concept of DT was introduced by NASA’s space missions in the early 2000s. During these missions, DTs created replicas of spacecraft systems on earth. This approach allowed ground teams to simulate and solve in-flight issues in real time \cite{glaessgen2012digital}. Since then, this idea has spread to manufacturing, driven by Industry 4.0’s focus on automation, data sharing, and interconnected smart systems \cite{kritzinger2018digitaltwin}.

DTs offer many advantages in manufacturing. For instance, DTs can leverage machine data for predictive maintenance to predict potential failures and make timely interventions to lessen downtime~\cite{qi2018digital}. DTs can also be used in quality control to replicate production lines or parts, allowing for digital inspections and measurement-based QC even before physical tests\cite{tao2018digitaltwin}. DTs also allow users to simulate changes in production environments within the virtual model and evaluate their outcomes without affecting real-life setups \cite{tao2018digitaltwin}.

By leveraging real-time simulations and feedback loops, DTs help prevalent analytical decision making that makes the best use of resources determines production cycles, furthers better product quality. All of this improved efficiency ultimately feeds into customer satisfaction, by reducing delays~\cite{kritzinger2018digital}. However, implementing DT technology in manufacturing brings specific challenges, particularly in predictive modeling and inspection, including:

\begin{itemize} \item \textbf{Data management and integration:} Manufacturing operations produce extensive sensor data, and incorporating this data accurately into DTs requires a robust setup that can handle high volumes, speeds, and types of data. Compatibility with older systems and diverse data formats also complicates integration \cite{tao2018digitaltwin, liu2022machining}.

\item \textbf{Predictive model accuracy:} DTs depend on precise, high-quality data from physical systems to accurately simulate performance. Incomplete or low-quality data can reduce predictive accuracy, limiting the effectiveness of maintenance forecasting. Effective models require comprehensive, well-organized datasets \cite{rosen2015autonomy, kritzinger2018digital}.

\item \textbf{Real-time synchronization and computational load:} Keeping physical and digital systems in real-time sync places demands on computational resources. Continuous updates can make it challenging to manage a large amount of data in real time and increase computational weights, reducing processing speed or inspection accuracy~\cite{glaessgen2012digital}.

\item \textbf{Security and privacy:} The continuous data flow between the physical and virtual environments can create cybersecurity risks in DT systems. A secure infrastructure is needed to protect manufacturing data and ensure ongoing monitoring~\cite{qi2018digital}.

\item \textbf{Integration with inspection processes:} While DTs improve inspection with simulations, aligning physical and virtual inspections is challenging. To effectively model a series of diverse manufacturing tasks, close integration with physical inspection is inevitable to ensure quality standards are maintained~\cite{zhang2023substation}. \end{itemize}

\subsection{Motivation of the Study}

In the era of Industry 4.0, the growing complexity and increasing precision required for manufacturing made advanced methods such as DTs inevitable. This study underscores the importance of employing DTs in quality control and failure prediction to minimize risks while ensuring reliable operations, especially in critical environments.

\begin{figure}[h]
    \centering
    \includegraphics[width=0.8\linewidth]{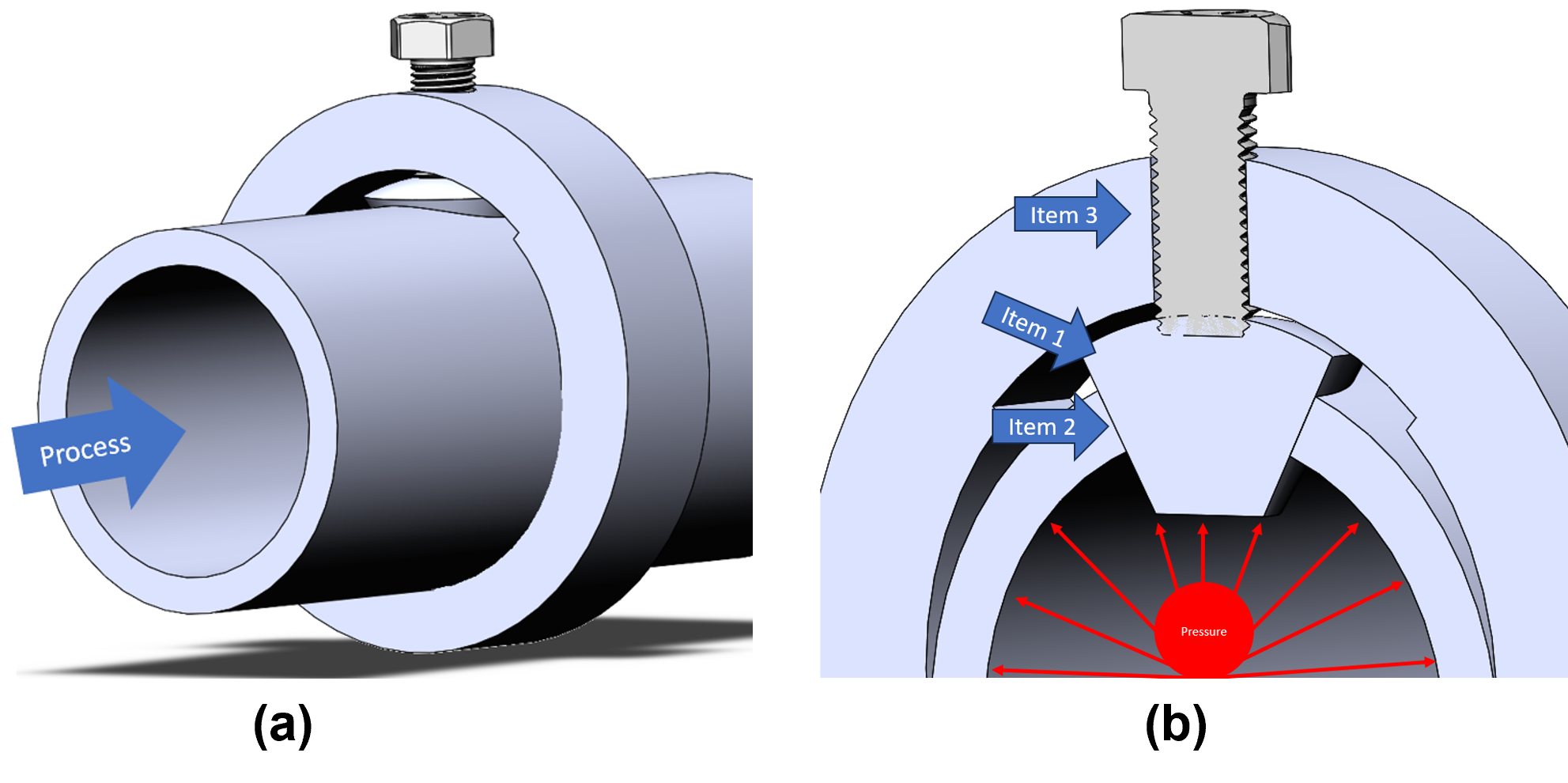}
    \caption{Conical hole plug assembly. (a) An external view of the process flow in the conical hole plug assembly. (b) A detailed view of the plug mechanism. Item 1 represents the conical plug, Item 2 is the conical hole, and Item 3 is the 1-1/2”–6 Grade 8 bolt. The red arrows depict internal pressure at 1400 bar (22,000 psi). }
    \label{fig:plug_assembly}
\end{figure}

For instance, a notable manufacturing challenge involved a conical hole plug assembly that failed, resulting in a dangerous projectile in a manufacturing environment. The conical plug, shown in Figure~\ref{fig:plug_assembly}(a), was designed to fit within a conical hole. Compressed by a 1-1/2”–6 Grade 8 bolt (Figure~\ref{fig:plug_assembly}(b)), it formed a seal to manage elastomer extrusion under high pressure (up to 1400 bar/22,000 psi). Over time, repeated use caused the bolt threads to fatigue and deform, ultimately leading to failure. A predictive maintenance system based on DT technology could have foreseen this failure through early detection, and lifecycle analysis might help to prevent the incident.

\begin{figure}[h]
    \centering
    \includegraphics[width=0.4\linewidth]{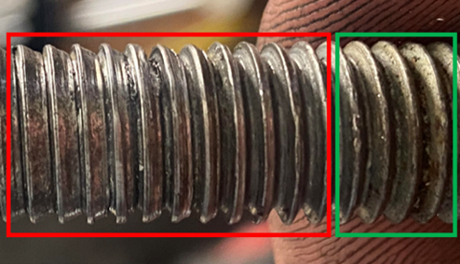}
    \caption{Deformed vs acceptable threads of the Bolt.}
    \label{fig:bolt_failure}
\end{figure}

During the elastomer extrusion process, extreme pressure on the conical hole plug exerted an upward force on the bolt threads. This repeated stress across multiple cycles deformed the threads, as shown in Figure~\ref{fig:bolt_failure}. The deformation eventually led to bolt failure and the release of uncontrolled pressure. A robust thread inspection process combined with real-time monitoring could have predicted this failure and enabled preventive maintenance to ensure safety.

Considering the opportunity, this study aims to develop integrated DT approaches to improve quality control and prevent failures in manufacturing processes through the proposed Predictive Maintenance and Inspection DT (PMI-DT) framework. The objectives of this study are outlined below:

\begin{enumerate}
    \item Create a DT using SolidWorks and Azure DTs to represent the CyberGage 360 and a 1”-4 ACME bolt, with model schema based on 3D mechanical models and critical dimensions.

    \item Design a workflow and data stream to validate the CyberGage 360’s inspection process and integrate fatigue testing and dimensional inspection data into the Bolt DT, forming the basis of the PMI-DT framework.

    \item Apply Machine Learning (ML) algorithms—Random Forest and Decision Tree models—to analyze inspection data and fatigue test results to detect patterns that indicate component failure.
\end{enumerate}

\section{Literature Review}
DT technology has gained significant attention in manufacturing as it provides advantages in process optimization, real-time monitoring, quality control, and product lifecycle management. Research identifies both the advantages and challenges of implementing DTs in these areas, with a growing interest in their role in improving quality control and preventing system failures.

One area of focus is process optimization. For instance, Yang et al. (2024) developed a DT-based solution in solid-wood-panel processing, where DTs combined with dynamic scheduling algorithms increased efficiency by 8.93\%~\cite{yang2024manufacturing}. Raudenbush et al. (2024) applied DTs in pharmaceutical manufacturing to improve processes by monitoring metabolite concentrations in real time and demonstrated DTs' flexibility in supporting ongoing improvements across industries~\cite{raudenbush2024pharmaceutical}.

DT technology supports advanced quality control with real-time oversight of manufacturing processes, allowing early detection of potential issues—an essential function in high-risk manufacturing settings. Real-time DT increases transparency and enables prompt issue resolution in wood-panel production~\cite{yang2024manufacturing}. In the pharmaceutical sector, DTs improve product quality by enabling constant adjustments to bioreactor conditions, showing their effectiveness in maintaining consistent quality standards across diverse production environments~\cite{raudenbush2024pharmaceutical}.

DTs improve product lifecycle management (PLM) by providing full visibility from design to disposal. Pronost et al. (2024) analyzed DT applications across PLM stages and noted that DTs most often support production and operational phases through real-time observation and predictive maintenance. However, gaps appear in DT use during design and disposal phases, where predictive functions could extend product life and improve end-of-life management~\cite{pronost2024product}. These results point to areas for further research in lifecycle management.

Integrating DTs with ML proves effective for predictive maintenance and fault detection in manufacturing. Patel and Kalgutkar (2024) applied a Random Forest model to predict machine states in industrial motors, which optimized maintenance through sensor data analysis~\cite{patel2024maintenance}. Their approach shows that ML models, when combined with DTs, manage large datasets to refine maintenance schedules. Similarly, Gitundu Kairo (2024) demonstrated that Deep Learning models, like Neural Networks, achieve high accuracy in failure prediction, further improved by ensemble methods~\cite{gitundu2024maintenance}. A study conducted by Smaoui and Baklouti (2024) compared multiple ML algorithms for fault detection; they found that calibrated K-Nearest Neighbors models obtained the highest accuracy of sensor-based machinery~\cite{smaoui2024fault}. Both of these cases illustrate how the integration between DT and ML can significantly benefit proactive maintenance, minimizing unplanned downtime and other operational inconsistencies that may occur.

IoT and ML technologies, as part of DTs, further improve predictive maintenance. Harsh et al., in 2024, demonstrated that IoT-based sensors predict failure rates using ML algorithms by analyzing conditions, e.g., temperature and vibration. It transforms maintenance from reactive to predictive and ultimately improves operational efficiency. This use of real-time data collection and predictive insights helps with advanced monitoring to ensure a safe, controlled environment in manufacturing industries~\cite{harsh2024iot}.

DTs also enhance quality control to serve as a complementary solution with more advanced inspection tools that assist in fulfilling accurate and faster inspections. For instance, a study conducted by Matthews et al. (2022) used the CyberGage 360, a highly automated robot-mounted 3D scanner for rapid and precise coverage measurement that reduces time and increases efficiency in quality control. DT systems also support tools such as laser scanning and photogrammetry. Put them together, and they make DTs for offshore infrastructure that are much more accurate, helping with remote inspections and streamlining the workflow. It allows more accurate identification of failures, makes it easier to perform quality checks, and lowers the cost~\cite{matthews2022inspection}.

In summary, based on the existing literature, DTs play a crucial role in process improvement, real-time tracking, and lifecycle management in manufacturing. However, the lack of sufficient research on quality control and failure prevention during the early and final stages of the lifecycle indicates the need for additional research on DT in AM to understand DT applications on a broader scale. Current studies confirm DTs’ usefulness across industries, but deeper integration with predictive ML models and inspection technologies could further help maintain quality and reduce failure risks.

\section{Methodology}
\begin{figure}[h]
    \centering
    \includegraphics[width=\linewidth]{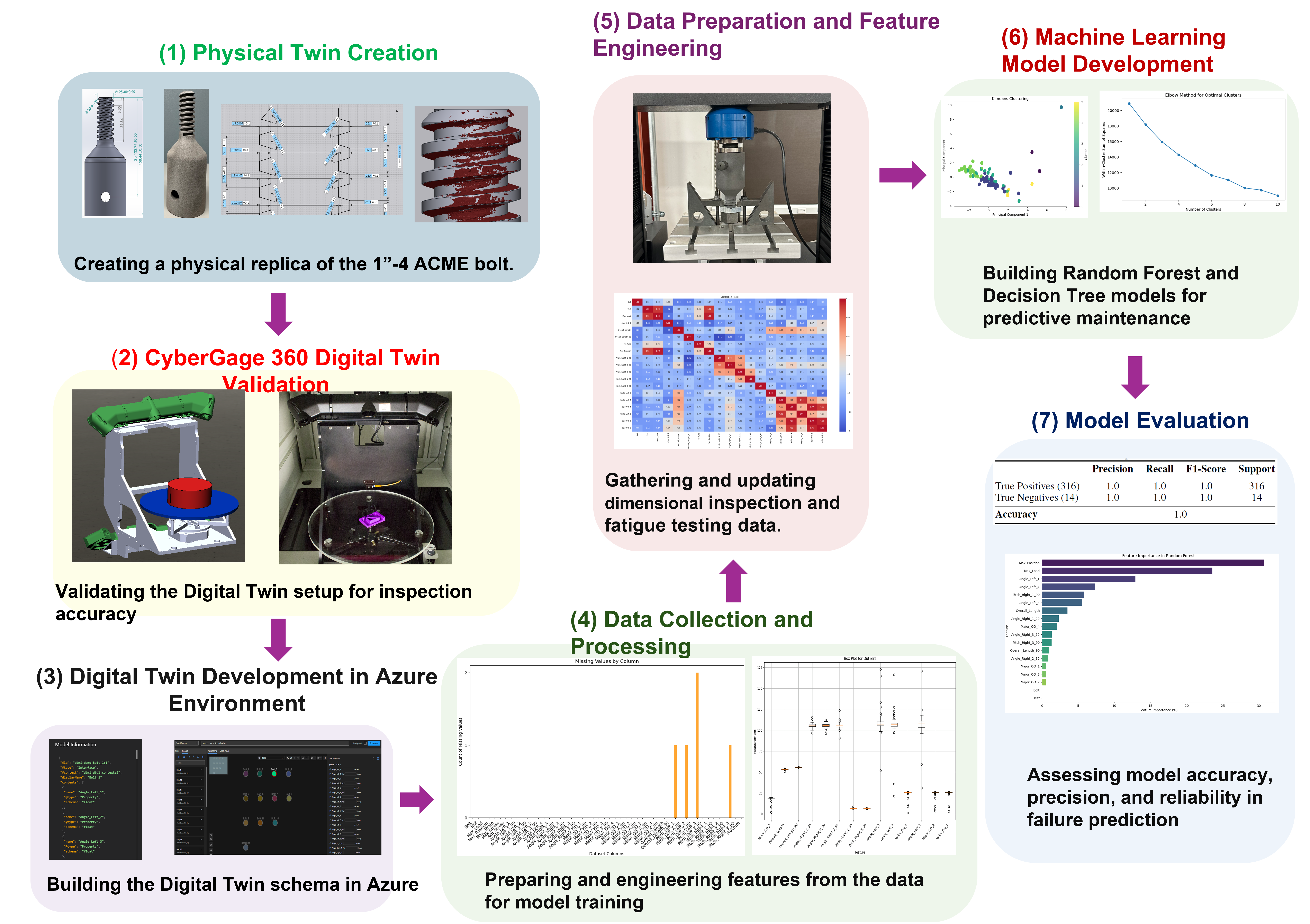}
    \caption{Proposed framework of Predictive Maintenance and Inspection DT (PMI-DT) for ACME bolts, including seven stages: (1) Physical Twin Creation, (2) CyberGage 360 DT Validation, (3) DT Development in the Azure Environment, (4) Data Collection and Processing, (5) Data Preparation and Feature Engineering, (6) Machine Learning Model Development, and (7) Model Evaluation.}
    \label{fig:PMIDT}
\end{figure}

Our proposed approach consists of several steps, which include: Physical Twin Creation, CyberGage 360 DT Validation, DT Development in the Azure Environment, Data Collection and Processing, Data Preparation and Feature Engineering, Machine Learning Model Development, and Model Evaluation, as shown in Figure~\ref{fig:PMIDT}. Each step is discussed in detail in the following subsections.
\subsection{Physical Twin Creation}

\begin{figure}[h]
    \centering
    \includegraphics[width=0.4\textwidth]{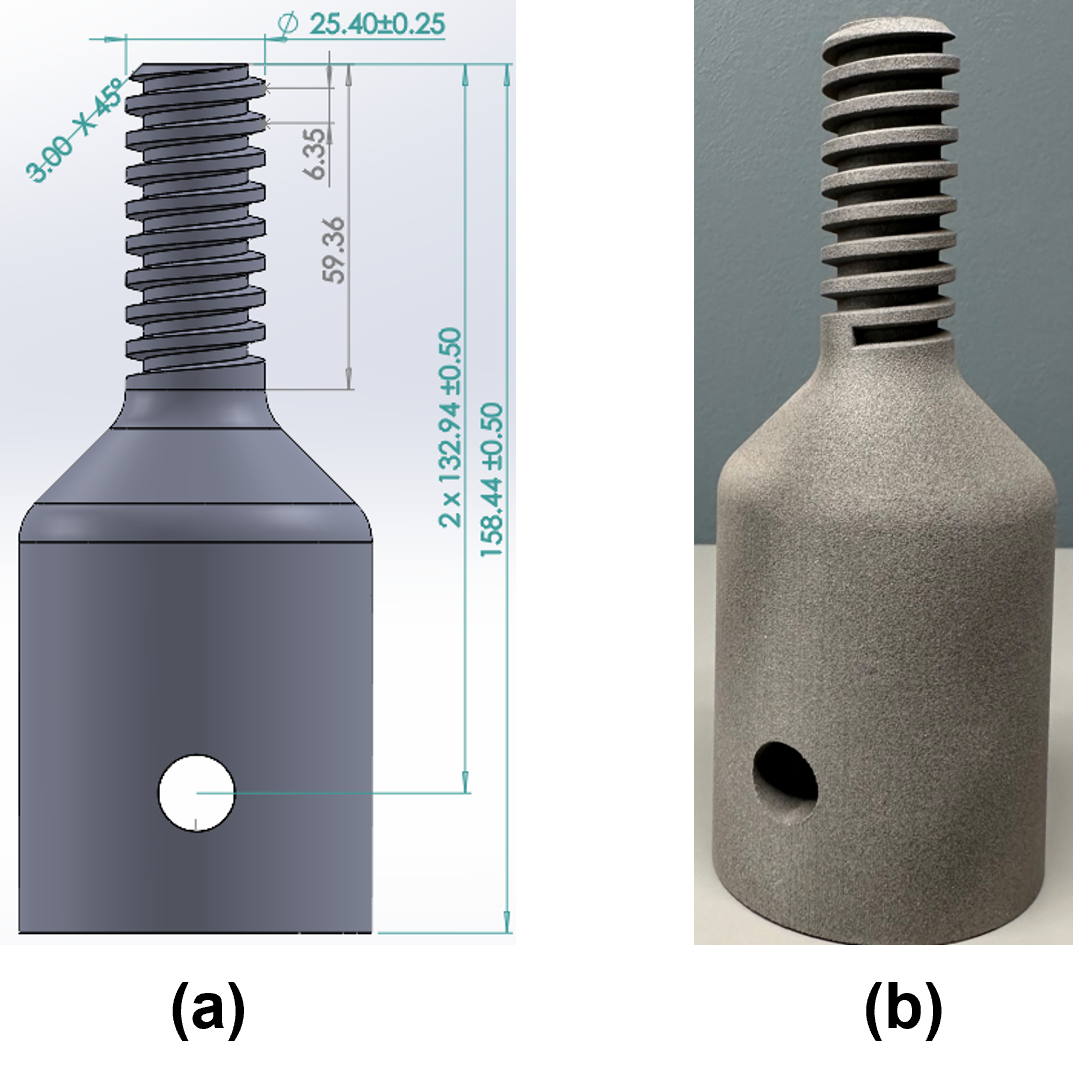}
    \caption{Illustration of (a) the fully defined 3D CAD model of the 1”-4 ACME Bolt and (b) the 3D-printed physical twin of the 1”-4 ACME Bolt using Nylon-12.}
    \label{fig:bolt_3dmodel}
\end{figure}

The Physical Twin of the 1”-4 ACME Bolt was created using Selective Laser Sintering (SLS) methods. The bolt design started in 3D CAD using SolidWorks, where a fully defined model was generated, and critical features were labeled using model-based definitions. The material chosen for the Physical Twin was powdered Nylon-12, selected for its high tensile strength (7,252 psi) and brittle behavior, as it has an elongation at break of approximately 4\%. Figure~\ref{fig:bolt_3dmodel}(a) shows the CAD model used for the Physical Twin, while Figure~\ref{fig:bolt_3dmodel}(b) depicts the 3D printed bolt used for testing.

\begin{figure}[h]
    \centering
    \includegraphics[width=0.7\textwidth]{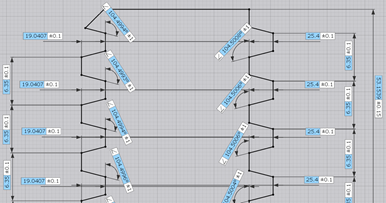}
    \caption{Nominal critical features identified in Geomagic Control X for the ACME Bolt.}
    \label{fig:bolt_criticalfeatures}
\end{figure}

\begin{figure}[h]
    \centering
    \includegraphics[width=0.3\textwidth]{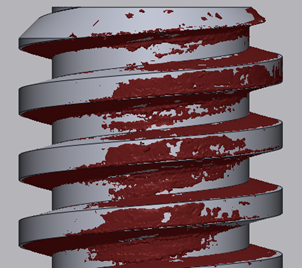}
    \caption{Heatmap comparison between 3D scanned model and theoretical CAD model.}
    \label{fig:bolt_heatmap}
\end{figure}

Once the Physical Twin was printed, baseline dimensional data were collected using the Geomagic Control X software. This software allows comparison between the laser-scanned model and the theoretical CAD model to visualize deviations and identify discrepancies. The critical features, such as thread angles and diameters, were cross-checked, as shown in Figure~\ref{fig:bolt_criticalfeatures}. Additionally, a heatmap analysis of nominal dimensions and scanned data provides valuable insights into potential deformations or inconsistencies in manufacturing (Figure~\ref{fig:bolt_heatmap}).

The Physical Twin was then subjected to tensile testing, where key data like maximum force and elongation were captured. This data formed the baseline for comparison with the DT in the later stages of the process.
\subsection{CyberGage 360 DT Validation}

Before ingesting data into the DT of the ACME Bolt, the CyberGage 360 must be validated to ensure that it can perform a complete dimensional inspection. The CyberGage 360 utilizes six image sensors, two of which are dual-camera optical blue light scanning sensors. These sensors are mounted above and below the subject part, which is placed on an optically flat glass plate, calibrated for scanning. The glass plate permits data capture from both sensors at once, which removes the need to flip the part as other systems require.

\begin{figure}[h]
    \centering
    \includegraphics[width=0.6\textwidth]{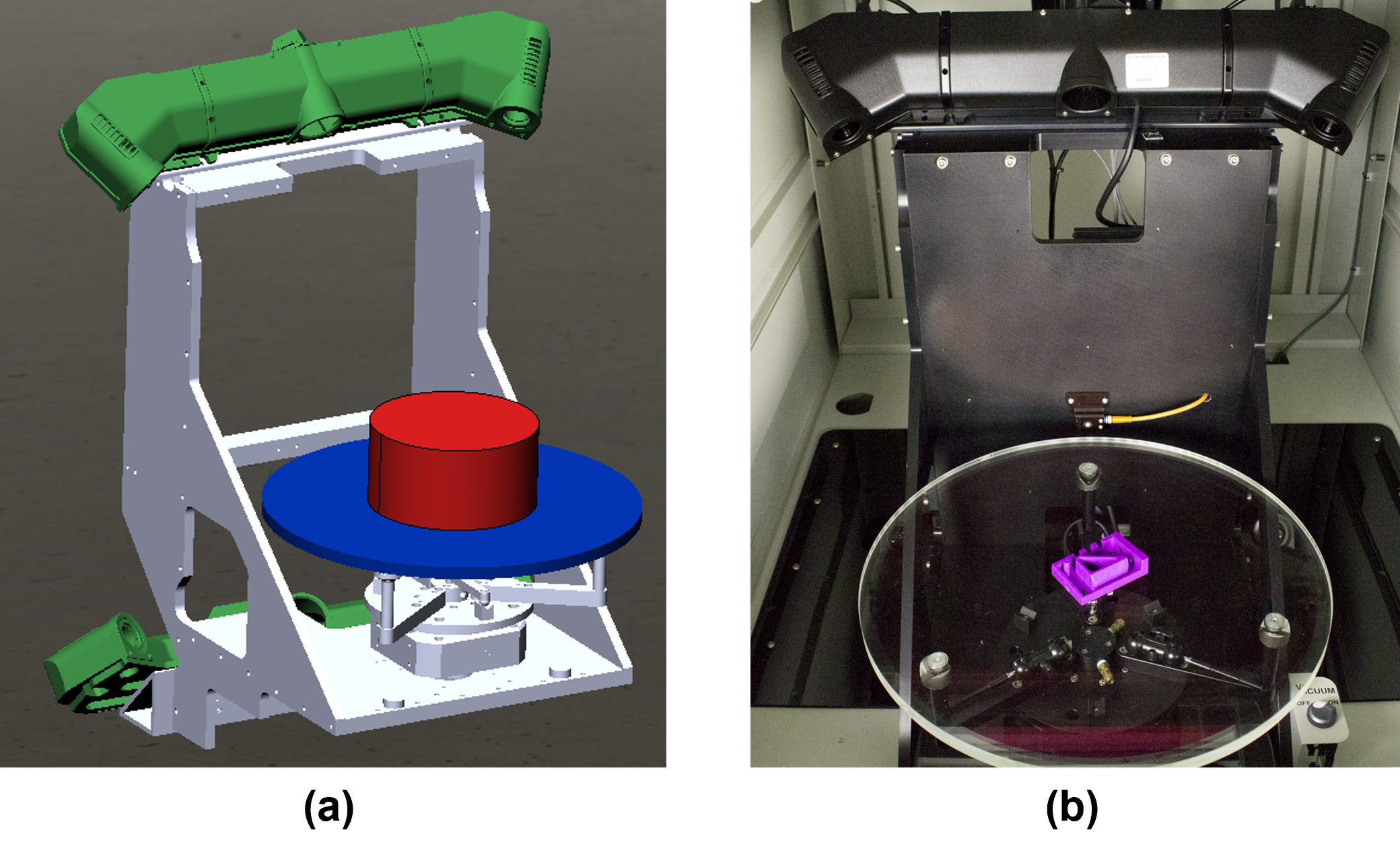}
    \caption{(a) CyberGage 360 DT and (b) CyberGage 360 Physical Twin.}
    \label{fig:cybergage_twin}
\end{figure}

The inspection area of the CyberGage 360 consists of a cylindrical space (200 mm diameter × 100 mm height), and the part must fit entirely within this space to ensure full inspection coverage. Figure~\ref{fig:cybergage_twin} (a) shows an overview of the CyberGage 360 DT, where green assemblies represent the six image sensors, the blue plate represents the rotating inspection plate, and the red cylinder represents the cylindrical inspection area, while Figure~\ref{fig:cybergage_twin} (b) illustrates the Physical Twin of the CyberGage 360.

\begin{figure}[h]
    \centering
    \includegraphics[width=0.8\textwidth]{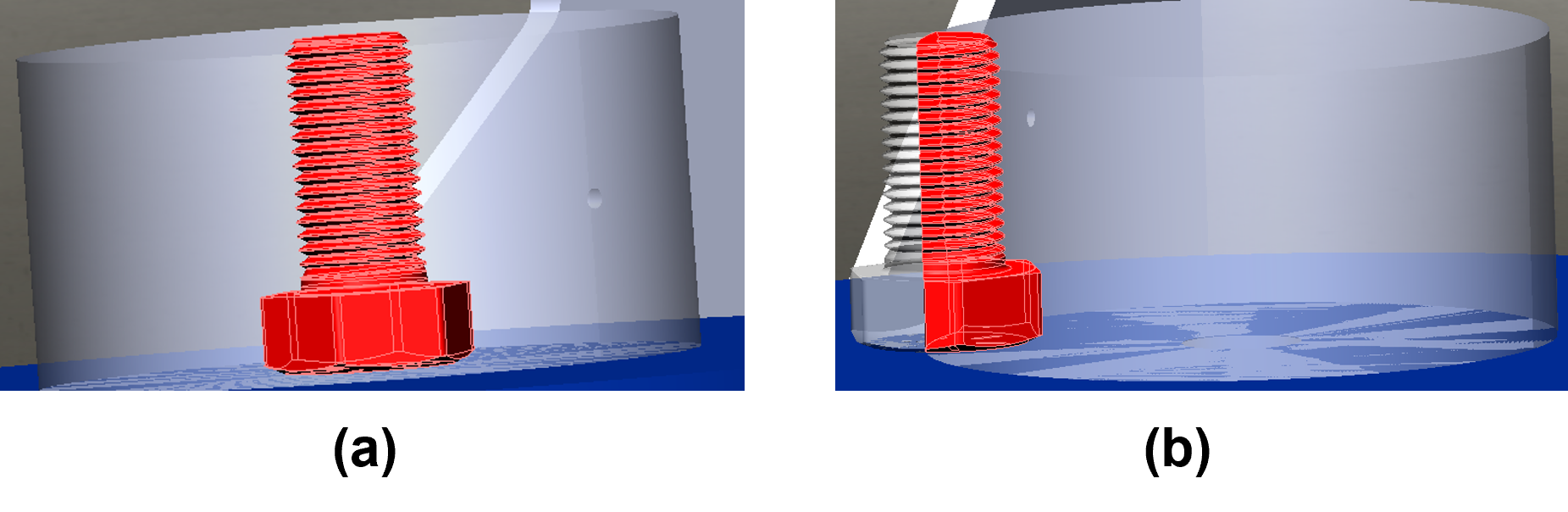}
    \caption{ACME Bolt positioned (a) inside the inspection region (red) and (b) outside the inspection region (silver).}
    \label{fig:bolt_inside}
\end{figure}

To validate that the ACME Bolt Physical Twin fits entirely within the inspection region, an interference detection process was performed between the bolt and the inspection area. Figure~\ref{fig:bolt_inside} (a) shows the bolt fully enclosed within the inspection area, with the red portion indicating correct positioning within the cylindrical space, while Figure~\ref{fig:bolt_inside} (b) illustrates a case where the bolt sits outside the inspection region due to incorrect placement, marked by the silver portion indicating the out-of-bounds section.

\begin{figure}[h]
    \centering
    \includegraphics[width=0.46\textwidth]{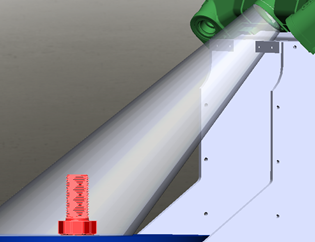}
    \caption{Interference detection between the light cone from the image sensor and the ACME Bolt.}
    \label{fig:light_cone}
\end{figure}

After confirming that the ACME Bolt fits within the inspection region, the next step is to ensure that all critical features of the bolt are visible to the six image sensors. This is achieved through a “light cone” analysis, where interference between the light cone and the part is detected to identify hidden or shadowed areas that the sensors cannot capture. Figure~\ref{fig:light_cone} shows the interference detection process, where shadows are identified to ensure complete inspection.

Once the CyberGage 360 DT is validated, the Physical Twin of the ACME Bolt can undergo full-dimensional inspection. The inspection data will then be ingested into the DT environment for further analysis and predictive maintenance simulations, as described in the subsequent sections.
\subsection{DT Development in the Azure Environment}

Following the validation of the CyberGage 360 DT, the next step involves creating a DT model for the ACME Bolt within the Azure DT environment. The DT of the bolt leverages data from both the Physical Twin and sensor information gathered during inspections. Azure DT’s environment enables structured input using DT Definition Language (DTDL), which helps define and manage various properties and parameters of the digital model.

To create the DT model, a JSON file defines the bolt's schema. Table~\ref{tab:bolt_schema} shows the primary attributes in the DT schema, with properties such as overall length, major diameter, angle, maximum load, maximum position, and fracture status. These properties allow a comprehensive digital representation of the bolt that can be used for predictive analysis.

\begin{table}[h]
    \centering
    \caption{Bolt DT schema example.}
    \label{tab:bolt_schema}
    \begin{tabular}{@{}lll@{}}
        \toprule
        \textbf{Name} & \textbf{@Type} & \textbf{Schema} \\
        \midrule
        Overall\_Length & Property & Float \\
        Major\_Diameter\_1 & Property & Float \\
        Angle\_Left\_1 & Property & Float \\
        Max\_Load & Property & Float \\
        Max\_Position & Property & Float \\
        Fracture & Property & Boolean \\
        \bottomrule
    \end{tabular}
\end{table}

\begin{figure}[h]
    \centering
    \includegraphics[width=0.3\textwidth]{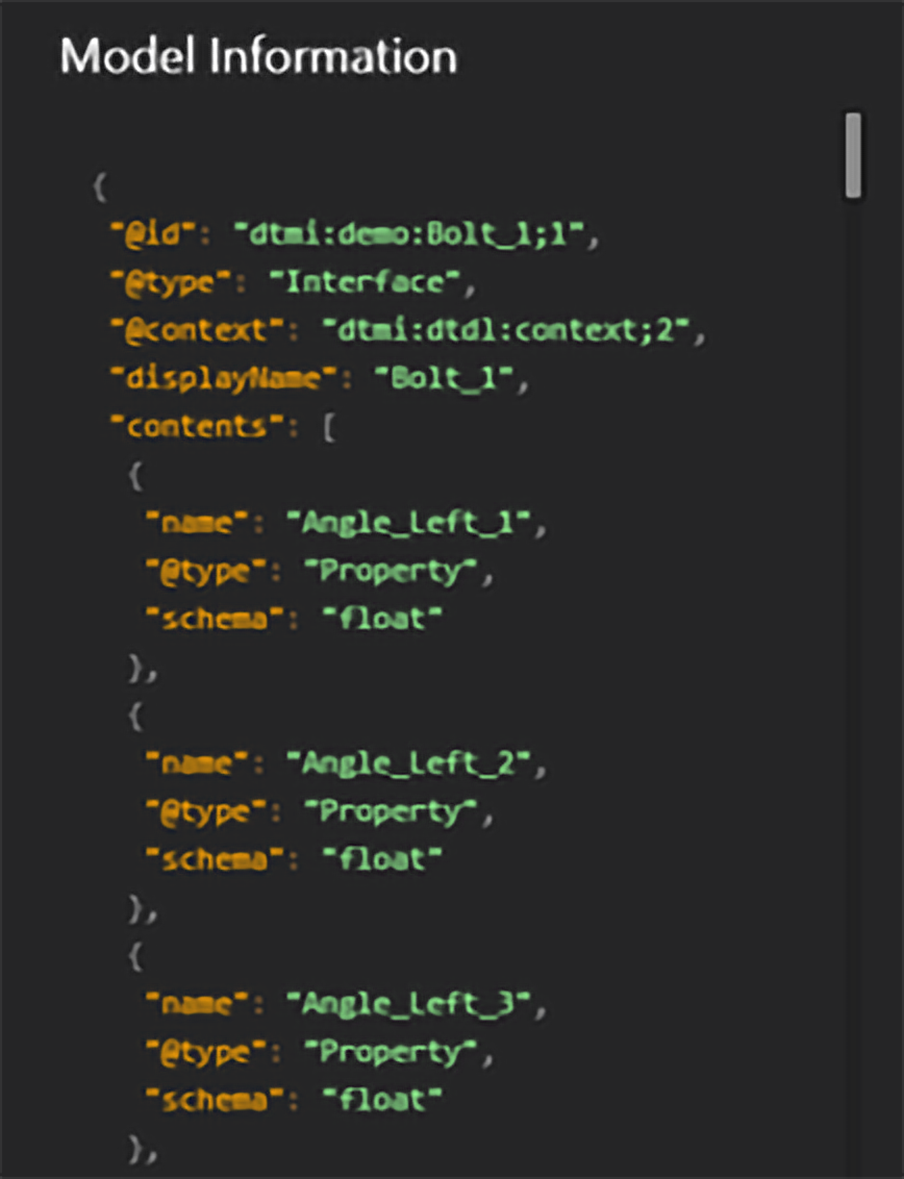}
    \caption{DT schema configuration in Azure environment.}
    \label{fig:azure_schema}
\end{figure}

Within the Azure DT platform, the DT model is created for multiple instances of the ACME Bolt, representing multiple Physical Twins. Each instance is initialized with preliminary inspection data obtained from the CyberGage 360, which enables a tailored and accurate virtual representation for each bolt. Figure~\ref{fig:azure_schema} illustrates the DT schema as visualized in Azure, highlighting the input parameters and their configurations.

\begin{figure}[h]
    \centering
    \includegraphics[width=0.8\textwidth]{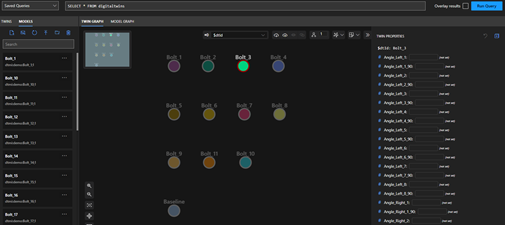}
    \caption{Initialized DT properties for ACME Bolt \#3 in Azure DT.}
    \label{fig:bolt_data}
\end{figure}

For this project, 10 DTs are generated, each representing one ACME Bolt Physical Twin. The inspection data from the CyberGage 360 is mapped onto each DT, facilitating a high level of accuracy in the digital replication. An example of the initialized data for Bolt \#3 is presented in Figure~\ref{fig:bolt_data}, showing critical parameters loaded into the Azure environment.

\subsection{Data Collection and Processing}

With the DT of the 1”-4 ACME Bolt generated, data collection begins by performing tensile testing on the 10 Physical Twins of the bolt. The test plan, outlined in Table~\ref{tab:test_plan} specifies the tensile elongation each bolt undergoes per cycle, applied for at least 10 cycles or until bolt failure. Each cycle records data in real-time, capturing both tensile test metrics and dimensional inspection results, which the system later transfers to the respective DT in the Azure environment.

\begin{table}[h!]
    \centering
    \caption{Tensile testing results for each bolt across multiple tests.}
    \label{tab:test_plan}
    \begin{tabular}{@{}l c c c c c c c c c c@{}}
        \toprule
        & \textbf{Bolt 1} & \textbf{Bolt 2} & \textbf{Bolt 3} & \textbf{Bolt 4} & \textbf{Bolt 5} & \textbf{Bolt 6} & \textbf{Bolt 7} & \textbf{Bolt 8} & \textbf{Bolt 9} & \textbf{Bolt 10} \\
        \midrule
        \textbf{Test 1} & 0.055 & 0.055 & 0.052 & 0.051 & 0.047 & 0.05 & 0.05 & 0.045 & 0.055 & 0.05 \\
        \textbf{Test 2} & \textcolor{red}{Failure} & 0.04 & 0.049 & 0.046 & 0.049 & 0.051 & 0.047 & 0.045 & 0.047 & 0.052 \\
        \textbf{Test 3} & \textcolor{red}{Failure} & \textcolor{red}{Failure} & 0.044 & 0.052 & 0.047 & 0.041 & 0.051 & 0.057 & 0.04 & 0.049 \\
        \textbf{Test 4} & \textcolor{red}{Failure} & \textcolor{red}{Failure} & 0.045 & 0.049 & 0.053 & 0.055 & 0.058 & 0.041 & 0.044 & 0.047 \\
        \textbf{Test 5} & \textcolor{red}{Failure} & \textcolor{red}{Failure} & 0.043 & 0.055 & 0.059 & 0.045 & 0.054 & 0.046 & 0.053 & 0.054 \\
        \textbf{Test 6} & \textcolor{red}{Failure} & \textcolor{red}{Failure} & 0.063 & 0.062 & 0.061 & 0.06 & 0.064 & 0.064 & 0.063 & 0.054 \\
        \textbf{Test 7} & \textcolor{red}{Failure} & \textcolor{red}{Failure} & 0.064 & 0.065 & 0.062 & 0.063 & 0.067 & 0.068 & 0.067 & 0.062 \\
        \textbf{Test 8} & \textcolor{red}{Failure} & \textcolor{red}{Failure} & 0.064 & 0.07 & 0.063 & 0.069 & 0.065 & 0.064 & 0.063 & 0.072 \\
        \textbf{Test 9} & \textcolor{red}{Failure} & \textcolor{red}{Failure} & 0.068 & 0.064 & 0.066 & 0.07 & 0.068 & 0.075 & 0.066 & 0.072 \\
        \textbf{Test 10} & \textcolor{red}{Failure} & \textcolor{red}{Failure} & 0.074 & 0.76 & 0.076 & 0.74 & 0.078 & 0.079 & 0.075 & 0.078 \\
        \textbf{Test 11} & \textcolor{red}{Failure} & \textcolor{red}{Failure} & 0.08 & \textcolor{red}{Failure} & 0.081 & 0.081 & 0.083 & 0.084 & 0.083 & 0.082 \\
        \bottomrule
    \end{tabular}
\end{table}

\begin{figure}[h]
    \centering
    \includegraphics[width=0.45\textwidth]{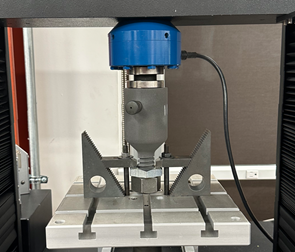}
    \caption{Physical Twin bolt setup for tensile testing.}
    \label{fig:bolt_tensile_setup}
\end{figure}

After each test cycle, the dataset is expanded with tensile test data and dimensional data for each bolt-test combination. This ensures that accurate and relevant data is available for ML model training, including tensile loads, elongation measurements, and the corresponding dimensional data. Figure~\ref{fig:bolt_tensile_setup} displays the physical setup for tensile testing of the bolts. After each cycle, this data enters the Azure DT platform, which provides an integrated digital model closely aligned with real-world conditions. This setup allows continuous updates and supports predictive maintenance and performance monitoring for each bolt.

\subsection{Data Preparation and Feature Engineering}

Once the data from tensile testing and dimensional inspection is collected and integrated into the Azure DT framework, it undergoes a series of preprocessing steps to ensure its suitability for ML modeling. These steps include handling missing values, performing feature engineering, and addressing outliers.

\subsubsection{Handling Missing Values}

\begin{figure}[h]
    \centering
    \includegraphics[width=0.8\textwidth]{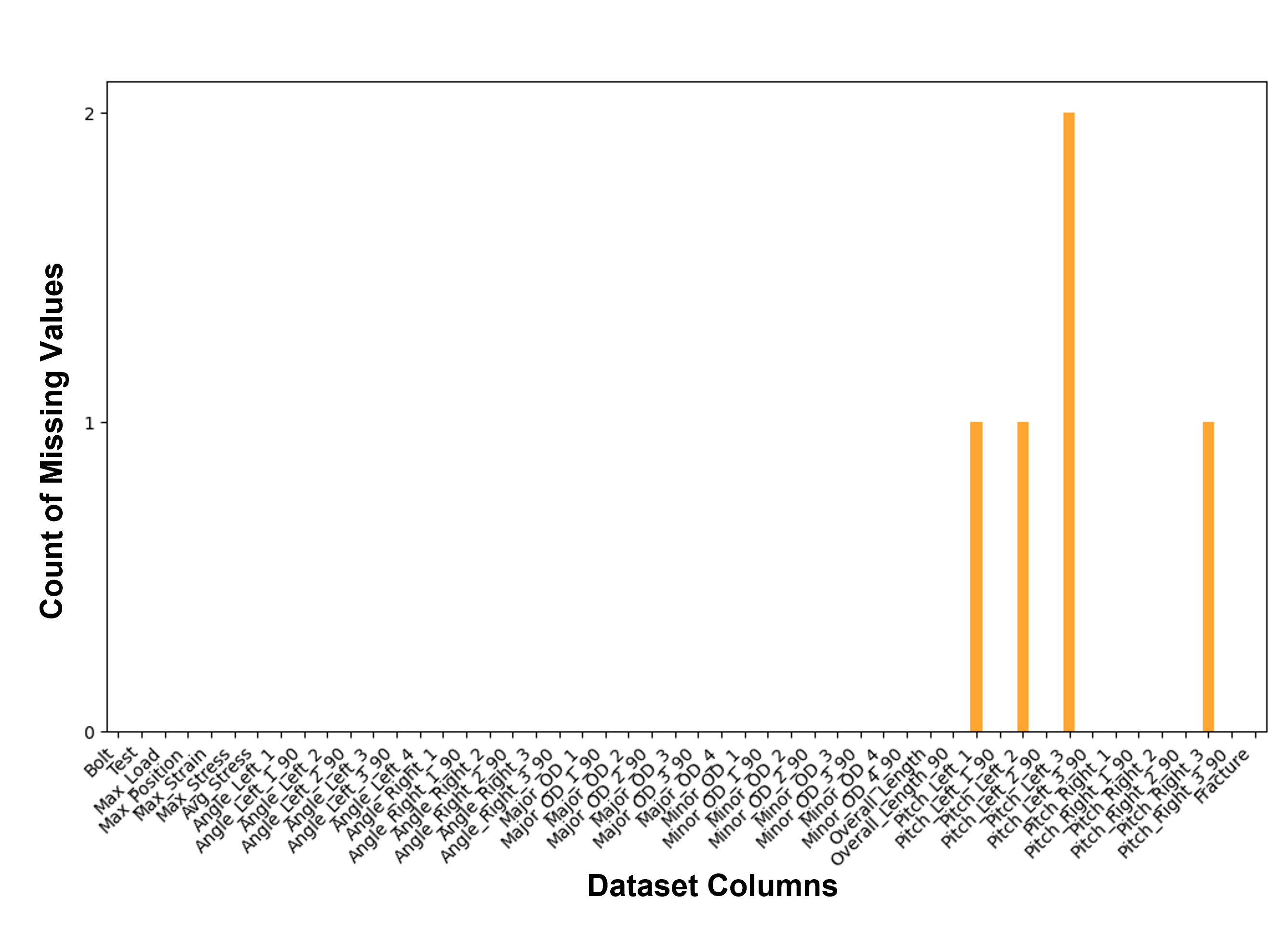} 
    \caption{Missing values within the dataset}
    \label{fig:missing_values}
\end{figure}

Despite automated data collection methods, minor data loss can occur. In this project, a preliminary analysis revealed
occasional missing values within the dimensional inspection data. As shown in Figure~\ref{fig:missing_values}, the dataset contains four columns with missing entries, each having one or two gaps. This results in approximately 5.8\% of the dataset’s 86 rows containing missing data.

Missing values appear exclusively within the dimensional inspection portion of the dataset, allowing a targeted approach to data imputation. Each missing entry replaces with a similar measurement taken at a 90-degree rotation for the same bolt and test combination. For instance, a missing value in the \texttt{Pitch\_Left\_1} field replaces with the \texttt{Pitch\_Left\_1\_90} measurement, recorded from an orthogonal angle. This approach ensures consistent data accuracy and preserves the statistical integrity of each bolt’s measurements.

\subsubsection{Feature Engineering}

To prepare the dataset for ML, relevant features are engineered from the raw data. Key features such as maximum force, maximum elongation, stress, and strain are calculated for each test. These features are important for predicting bolt failure, as they represent the material's response to tensile stress.

The calculations are performed using the following equations~\cite{h2005marks}:
\begin{equation}
    \sigma = \frac{F}{A}
\end{equation}
where \( F \) is the applied force, and \( A \) is the cross-sectional area of the bolt.

\begin{equation}
    \epsilon = \frac{\Delta L}{L_0}
\end{equation}
where \( \Delta L \) is the change in length, and \( L_0 \) is the initial length.

Additionally, a binary feature labeled as \texttt{Failure} is created to indicate whether a bolt failed during a particular test cycle. This serves as the target variable for predictive modeling.

\subsubsection{Outlier Detection and Handling}

\begin{figure}[h]
    \centering
    \includegraphics[width=0.8\textwidth]{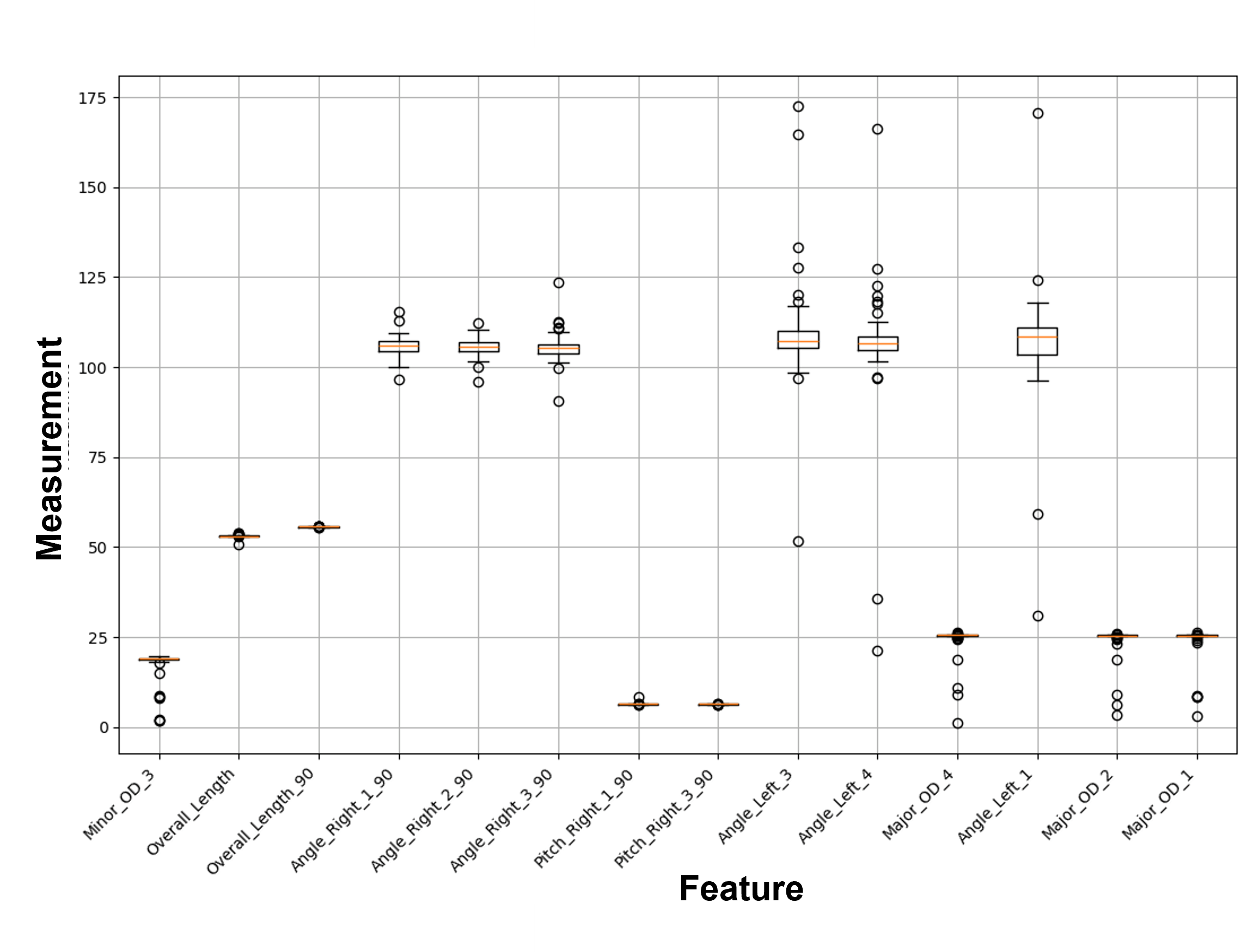} 
    \caption{Outlier values per feature.}
    \label{fig:outlier_values}
\end{figure}

The dataset is further refined by identifying and mitigating outliers. Outliers are detected using the z-score method, where any value with an absolute z-score above 3.0 is considered an outlier, excluding values in the target variable column, \textit{Fracture}. As illustrated in Figure~\ref{fig:outlier_values}, box plots for each feature column display the number of outliers per feature, highlighting values that exceed the z-score threshold.

For these outliers, values are imputed using the average from previous tests performed on the same bolt. This approach preserves the statistical integrity specific to each bolt’s history while minimizing the influence of outliers on the overall dataset.

Due to the limited sample size in this project, bootstrapping is applied to expand the dataset artificially. This technique involves creating additional samples by resampling with replacement from the original data. The dataset is extended to include 100 bolts, each with up to 11 tensile tests, resulting in a more robust training set for ML model deploment.

\subsection{Data Exploration}




To gain insights into the data structure, descriptive statistics were applied to evaluate the standard deviation of each feature. Table~\ref{tab:top5_stddev} shows the top five dimensional measurements with the highest standard deviations.

\begin{table}[h]
    \centering
    \caption{Top 5 dimensional measurements with highest standard deviation.}
    \label{tab:top5_stddev}
    \begin{tabular}{@{}ll@{}}
        \toprule
        \textbf{Feature} & \textbf{Standard Deviation (degrees)} \\
        \midrule
        Angle\_Right\_1 & 20.19 \\
        Angle\_Right\_2 & 19.73 \\
        Angle\_Right\_3 & 18.76 \\
        Angle\_Left\_4 & 14.24 \\
        Angle\_Left\_2 & 13.24 \\
        \bottomrule
    \end{tabular}
\end{table}

\begin{figure}[h]
    \centering
    \includegraphics[width=0.6\textwidth]{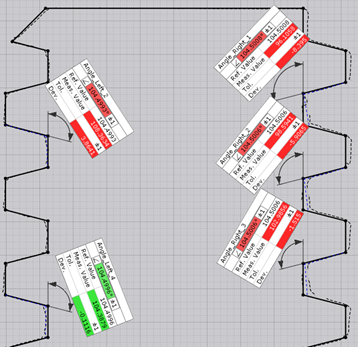}
    \caption{Bolt cross-section with dimensions of high standard deviation.}
    \label{fig:high_stddev_dimensions}
\end{figure}

As illustrated in Figure~\ref{fig:high_stddev_dimensions}, these five measurements, all related to thread angles, exhibit the highest standard deviation. These angles experience the maximum downward load applied by the tensile testing machine. While one might initially assume that the large standard deviations in these thread angles are due to deformation occurring during testing, supporting the project’s objectives, it remains uncertain if this is the actual cause. The observed deviation may also stem from limitations within the measurement system or inconsistencies in the 3D printing manufacturing process of the bolts, rather than as a direct result of tensile testing. Given that a standard ACME thread was used for the test specimens, where thread angles are highly controlled, further investigation would be necessary to confirm the origin of these variations.

\subsection{Model Selection and Training}

We have used Random Forest and Decision Tree algorithms to classify bolt failure using a dataset with high dimensionality. These two algorithms were chosen for their performance with high-dimensional data~\cite{ahsan2021effect}. 
The \texttt{train\_test\_split} function from Scikit-Learn was used to divide data into training and testing sets, with the \textit{Fracture} feature as the target variable, representing bolt failure. The dataset included 1,100 samples, with 9\% (100 samples) indicating failure cases. To ensure reproducibility, a \texttt{random\_state} parameter was set. A 70\%-30\% split resulted in:
\begin{itemize}
    \item \textbf{Training set size}: $0.7 \times 1100 = 770$ samples
    \item \textbf{Testing set size}: $0.3 \times 1100 = 330$ samples
\end{itemize}

\subsection{Machine Learning Models}
\textbf{Model One: Random Forests}

The Random Forest algorithm is an ensemble learning method that builds multiple decision trees and combines them to improve prediction accuracy and control overfitting. Each tree in a Random Forest is trained on a random subset of the training data and features, which enables diversity amongst the trees. Each tree's prediction, on average, becomes the final prediction, which is done by majority vote in case of classification problems.

For a given input sample \( x \), each tree \( T_i \) in the forest predicts a class \( y_i \), and the final prediction \( y \) is made based on the majority vote:
\begin{equation}
y = \operatorname{mode}(T_1(x), T_2(x), \ldots, T_n(x))
\end{equation}
where \( n \) is the number of trees in the forest.

Each tree is constructed using a method called \textit{Bootstrap Aggregation} (or \textit{bagging}). We are sampling with replacement from the training data to create each tree and taking a random subset of features at every split. This decreases the correlation between trees and helps reduce overfitting.

The algorithm also uses \textit{Gini impurity} or \textit{entropy} to determine the best splits within each tree. For example, Gini impurity for a node is calculated as:
\begin{equation}
G = 1 - \sum_{i=1}^C p_i^2
\end{equation}
where \( p_i \) is the proportion of samples of class \( i \) in the node, and \( C \) is the total number of classes.

By aggregating the predictions of multiple uncorrelated trees, Random Forests achieve high accuracy and are robust against overfitting, especially in high-dimensional datasets~\cite{rigatti2017random}.

\textbf{Model Two: Decision Trees}

The Decision Tree algorithm is a supervised learning model that makes decisions by learning basic decision rules derived from the features in the training data. It recursively divides the dataset into subsets based on the feature that most effectively distinguishes the classes. Each internal node in the tree corresponds to a feature, each branch corresponds to a decision rule, and each leaf node corresponds to an outcome or prediction.

To determine the best split at each node, the algorithm uses criteria such as \textit{Gini impurity} or \textit{information gain}. For example, the Gini impurity, \( G \), for a node is defined as:
\begin{equation}
G = 1 - \sum_{i=1}^{C} p_i^2
\end{equation}
where \( p_i \) is the proportion of samples of class \( i \) in the node, and \( C \) is the total number of classes.

Alternatively, information gain, based on \textit{entropy}, can be used to measure the reduction in uncertainty:
\begin{equation}
\text{Entropy} = -\sum_{i=1}^{C} p_i \log_2(p_i)
\end{equation}
The \textit{information gain} for a split is calculated as:
\begin{equation}
\text{Information Gain} = \text{Entropy (parent)} - \sum_{k} \frac{|A_k|}{|A|} \text{Entropy (child}_k)
\end{equation}
where \( A_k \) are the child nodes resulting from the split, and \( |A| \) is the total number of samples.

The tree grows by repeatedly splitting nodes until a stopping criterion is met, such as a maximum depth or minimum number of samples per node, to prevent overfitting. In Decision Trees, the final prediction for a sample follows a path down the tree based on feature thresholds at each node, ending at a leaf node that represents the class prediction~\cite{song2015decision}.

\subsection{Evaluation Metrics}

Model performance was assessed through a Confusion Matrix, which evaluated True Positives (TP), False Positives (FP), True Negatives (TN), and False Negatives (FN). Based on this, a Classification Report detailed accuracy, precision, recall, and F1-Score. The evaluation metrics are calculated as follows~\cite{ahsan2024defect,ahsan2024enhancing}:

\begin{equation}
\text{Precision} = \frac{TP}{TP + FP}
\end{equation}

\begin{equation}
\text{Recall} = \frac{TP}{TP + FN}
\end{equation}

\begin{equation}
\text{F1-Score} = 2 \times \frac{\text{Precision} \times \text{Recall}}{\text{Precision} + \text{Recall}}
\end{equation}

where:

\begin{itemize}
    \item True Positive (\(t_p\)) = Instances correctly classified as positive.
    \item False Positive (\(f_p\)) = Instances incorrectly classified as positive.
    \item True Negative (\(t_n\)) = Instances correctly classified as negative.
    \item False Negative (\(f_n\)) = Instances incorrectly classified as negative.
\end{itemize}

\section{Results}

\subsection{Random Forest Model Performance}

Table~\ref{tab:rf_classification_report} presents the classification report of the Random Forest model's performance on the test set. The model achieved perfect scores in accuracy, precision, recall, and F1-score for both positive and negative classes, indicating accurate predictions of bolt failures.

\begin{table}[h]
    \centering
    \caption{Classification report for Random Forest model.}
    \label{tab:rf_classification_report}
    \begin{tabular}{@{}lcccc@{}}
        \toprule
        & \textbf{Precision} & \textbf{Recall} & \textbf{F1-Score} & \textbf{Support} \\
        \midrule
        True Positives (316) & 1.0 & 1.0 & 1.0 & 316 \\
        True Negatives (14) & 1.0 & 1.0 & 1.0 & 14 \\
        \midrule
        \textbf{Accuracy} & \multicolumn{4}{c}{1.0} \\
        \bottomrule
    \end{tabular}
\end{table}

The confusion matrix in Table~\ref{tab:rf_confusion_matrix} further confirms the model’s high performance, with all actual positives and negatives accurately classified, resulting in no False Positives or False Negatives.

\begin{table}[h]
    \centering
    \caption{Confusion matrix for Random Forest model.}
    \label{tab:rf_confusion_matrix}
    \begin{tabular}{@{}ccc@{}}
        \toprule
        & \textbf{Predicted Positive} & \textbf{Predicted Negative} \\
        \midrule
        \textbf{Actual Positive} & 316 & 0 \\
        \textbf{Actual Negative} & 0 & 14 \\
        \bottomrule
    \end{tabular}
\end{table}

\subsection{Decision Tree Model Performance}

Table~\ref{tab:dt_classification_report} presents the classification report for the Decision Tree model on the test set. The model achieved perfect scores across accuracy, precision, recall, and F1-score for both classes, indicating a high level of accuracy in identifying bolt failures.

\begin{table}[h]
    \centering
    \caption{Classification report for Decision Tree model.}
    \label{tab:dt_classification_report}
    \begin{tabular}{@{}lcccc@{}}
        \toprule
        & \textbf{Precision} & \textbf{Recall} & \textbf{F1-Score} & \textbf{Support} \\
        \midrule
        True Positives (316) & 1.0 & 1.0 & 1.0 & 316 \\
        True Negatives (14) & 1.0 & 1.0 & 1.0 & 14 \\
        \midrule
        \textbf{Accuracy} & \multicolumn{4}{c}{1.0} \\
        \bottomrule
    \end{tabular}
\end{table}

The confusion matrix in Table~\ref{tab:dt_confusion_matrix} supports the model's high accuracy, showing that all Actual Positives and Actual Negatives were correctly classified, with no False Positives or False Negatives.

\begin{figure}[h]
    \centering
    \includegraphics[width=\textwidth]{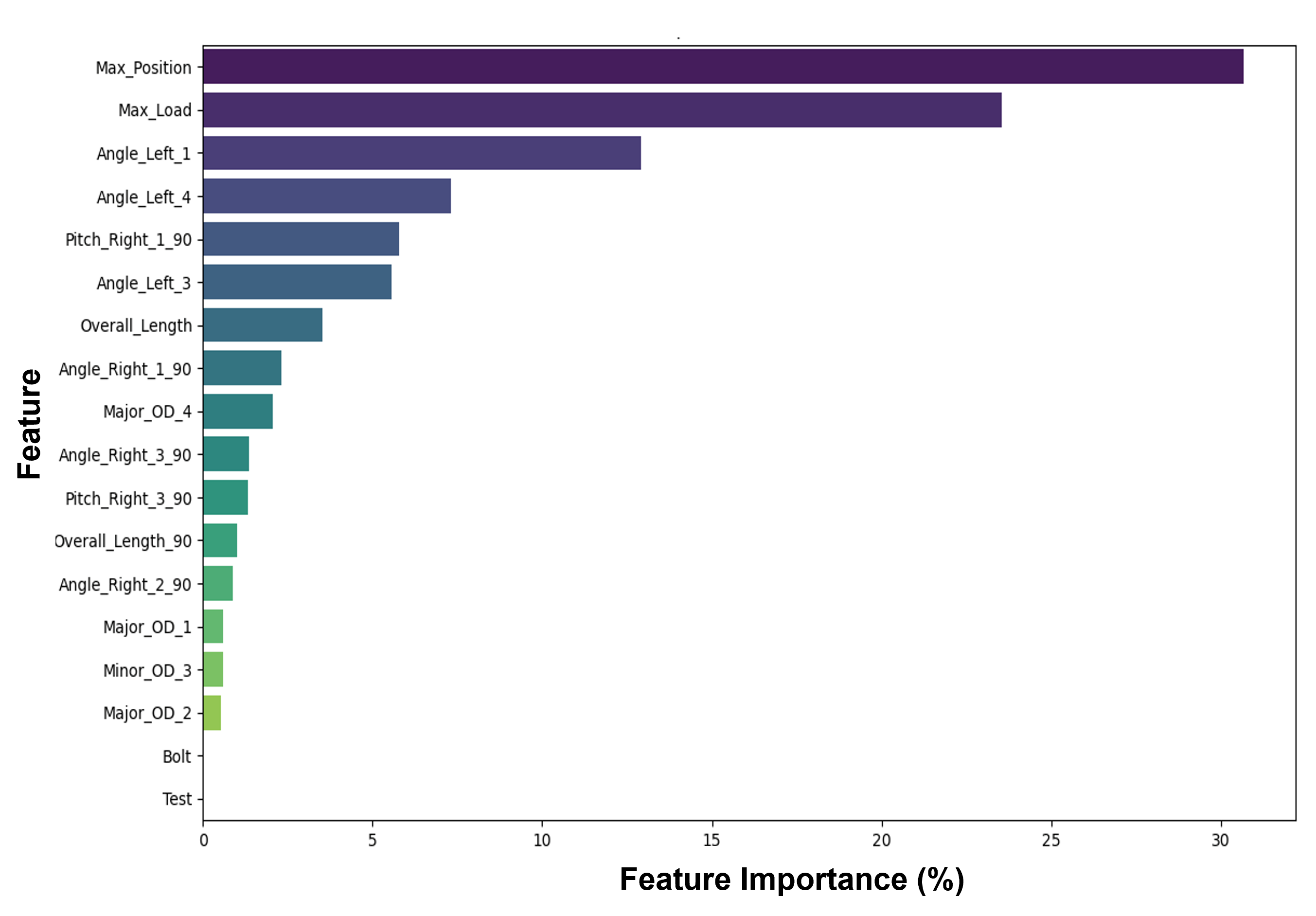}
    \caption{Feature importance in the Random Forest model.}
    \label{fig:feature_importance_rf}
\end{figure}

\begin{table}[h]
    \centering
    \caption{Confusion matrix for Decision Tree model.}
    \label{tab:dt_confusion_matrix}
    \begin{tabular}{@{}ccc@{}}
        \toprule
        & \textbf{Predicted Positive} & \textbf{Predicted Negative} \\
        \midrule
        \textbf{Actual Positive} & 316 & 0 \\
        \textbf{Actual Negative} & 0 & 14 \\
        \bottomrule
    \end{tabular}
\end{table}
To understand the critical features influencing model predictions, we analyzed feature importance using the \texttt{feature\_importances\_} attribute in Scikit-Learn. This analysis provides insight into the features that most significantly impact the classification results, guiding the interpretation of model outputs in relation to bolt failure prediction. The Random Forest model, by design, generates multiple decision trees and averages their outputs, allowing it to capture complex feature interactions effectively. According to the analysis of \texttt{feature\_importances\_}, the most impactful features were Max Position (30\%) and Max Load (24\%), with additional contributions from Left Thread Angle 1 (14\%) and Left Thread Angle 4 (8\%) (see Figure~\ref{fig:feature_importance_rf}). This distribution suggests that the model relies on both material strain (Max Position) and stress (Max Load) metrics, supported by additional dimensional features, for robust failure prediction.

\begin{figure}[h]
    \centering
    \includegraphics[width=\textwidth]{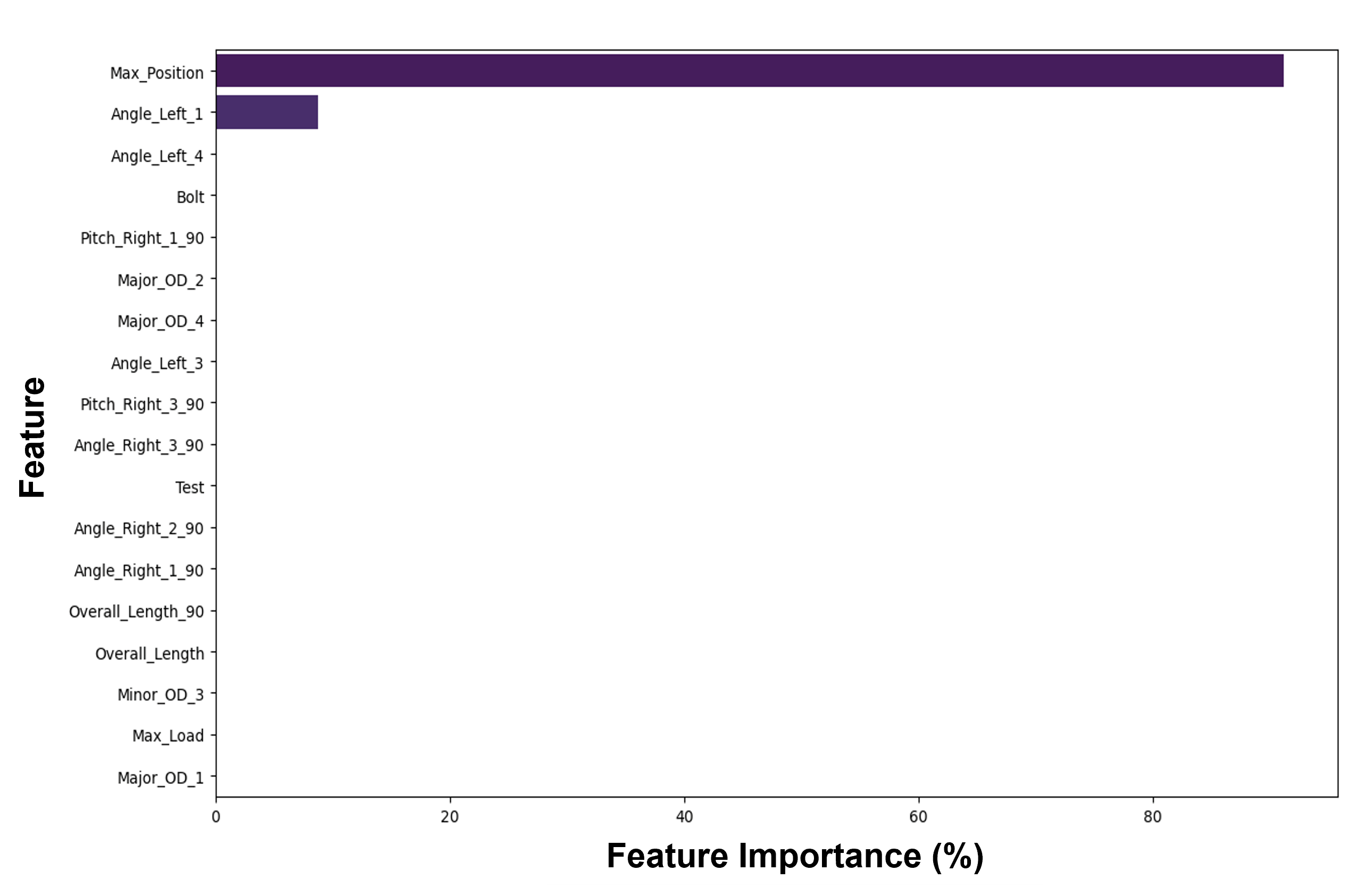}
    \caption{Feature importance in the Decision Tree model.}
    \label{fig:feature_importance_dt}
\end{figure}

Conversely, the Decision Tree model showed a much simpler structure in its feature importance analysis, with Max Position accounting for 95\% of the model's predictive power and Left Thread Angle 1 contributing only 15\% (Figure~\ref{fig:feature_importance_dt}). Given the Decision Tree’s single-branch structure, this result aligns with its tendency to focus heavily on dominant features, in this case, Max Position, for classification.

\section{Discussion}

In manufacturing, the ability to predict component failure is essential to minimize downtime and reduce maintenance costs. This study proposes the PMI-DT framework to predict bolt failures under tensile stress by simulating real-time scenarios. Using Random Forest and Decision Tree models, PMI-DT enables precise predictions, with both models achieving perfect classification scores, as detailed in Tables~\ref{tab:rf_classification_report}--~\ref{tab:dt_confusion_matrix}. The Random Forest model performs better due to its ensemble structure, which captures complex, non-linear relationships between features. Analysis identifies Max Position and Max Load as primary predictors of failure, aligning with the material properties of Nylon-12 bolts, whose brittleness under tensile stress demonstrates PMI-DT's predictive capacity.

However, Predictive accuracy, depends on the quality and frequency of data updates from physical systems to the DT environment. Delays or errors in synchronization may lead to inaccurate predictions, and reduce model reliability in real applications. Additionally, real-time data processing and model updates require high computational resources, which may limit scalability, especially in large manufacturing systems. Optimizing data pipelines and using distributed computing could help PMI-DT scale effectively and deliver timely predictions.

Integrating inspection data into PMI-DT provides a general overview of component status, which demonstrates PMI-DT’s capacity to identify wear and defects. During this study, we have also used the CyberGage 360 inspection tool to capture precise measurements of ACME bolts and map those measurement data onto the DT for predictive analysis. However, keeping all critical bolt features consistently visible and measurable also presents challenges. For instance, minor sensor misalignment can compromise data accuracy, particularly when capturing complex features like thread angles, which are essential for assessing bolt integrity under tensile loads.

Additionally, incorporating inspection data with manufacturing variability further complicates the integration process. For instance, even though the deviations in thread angles may reflect tensile stress effects, they might not be consistent throughout the printing process and ultimately affect the measured data. Therefore, future PMI-DT applications could incorporate multi-sensor fusion or cross-validation techniques that combine data from various inspection methods, such as laser scanning and photogrammetry, to ensure the measurement data is consistent throughout the printing process. 

During this research, the ACME bolts made of Nylon-12 showed good tensile strength but poor ductility, i.e., less plastic behavior in elongation prior to breaking under uniaxial tension. The ML-based feature analysis reveals that the Max Load and Max Position are the most essential features contributing to such failure. PMI-DT uses these properties to identify occurrence patterns (i.e., cyclic behavior related to brittle materials). However, the practicality of constitutive models based on those parameters for components made of more ductile material may be conditional. For example, for ductile materials, predictive models might have to account for strain hardening or residual stresses when loaded. Hence, tuning PMI-DT for each material-specific case matters to get accurate results. This includes modifying DT models to mimic the behavior of each material used for manufacturing.

\subsection{Limitations and Future Scope}
While this study demonstrates the potential of the PMI-DT framework in predictive maintenance, it also reveals several limitations and areas for future research:

\begin{itemize}
    \item \textbf{Limited dataset scope:} This study focuses on a single component type (ACME bolts) and one material (Nylon-12). To generalize the results, future studies should include a broader range of components and materials to assess how PMI-DT adapts to different mechanical properties and structural designs. A more diverse dataset would support comprehensive model training and validation, potentially improving predictive accuracy across varied manufacturing scenarios.

    \item \textbf{Model selection constraints:} This study relies on Random Forest and Decision Tree models, which, while effective, may not capture the full complexity of high-dimensional data in larger manufacturing systems. Future research could explore alternative ML models, such as Neural Networks or Gradient Boosting, to improve scalability and prediction accuracy. Hybrid DT models combining machine learning with physics-based simulations could also offer more detailed insights, especially for components with complex mechanical interactions.

    \item \textbf{Data integration and processing challenges:} Maintaining real-time synchronization between physical and digital systems presents computational challenges, particularly in high-volume environments. Future work should focus on developing efficient data pipelines and distributed processing frameworks to enable PMI-DT to scale effectively in industrial applications.
\end{itemize}

\section{Conclusion}

In this study, we have proposed an integrated DT framework, namely PMI-DT, to address the challenges of predictive maintenance and quality control in manufacturing environments. By integrating DT with ML models, specifically Random Forest and Decision Tree algorithms, we predicted the failure in manufacturing components, such as ACME bolts made from Nylon-12 100\% accurately. We have integrated advanced simulation and inspection methods in our proposed approaches, using tools such as CyberGage 360 for precise dimensional data collection and Microsoft Azure for DT management. We found that data from tensile testing and dimensional inspection can create a real-time feedback loop to help detect wear and potential failures. Early Our findings also suggest that features such as the Maximum Load and Maximum Position play key roles in bolt failure, mainly in materials with high brittleness. Once incorporated with ML, we found that the DT-based approaches can be a great asset that can save time and resources and improve downtime in the manufacturing world. We believe our proposed PMI-DT approaches will give some direction to future researchers and practitioners who want to explore the optimal opportunity of integrated DT and ML approaches in manufacturing domains. However, the proposed PMI-DT was not tested with a big data environment and different manufacturing settings, which we aim to explore in our future work. Apart from this, the future work will also try to explore several other opportunities, such as expanding the PMI-DT framework, which might support different components and materials, physics-based simulation with ML, multimodal and multisensory data with explainable AI, and also the limitation discussed in the discussion section.

\section*{Conflict of interest}
The authors declare no conflict of interest.
\bibliographystyle{unsrt}  
\bibliography{main}

\end{document}